\documentclass[10pt,conference]{IEEEtran}
\usepackage{subfigure}
\usepackage{amssymb}
\usepackage{amsthm}
\usepackage{amsmath}
\usepackage{hyperref}
\usepackage{graphicx}
\usepackage{colortbl,dcolumn}
\usepackage{booktabs}
\usepackage{xcolor}
\usepackage{cite}
\usepackage{threeparttable}

\usepackage{algorithm}
\usepackage{algorithmic}

\newtheorem{theorem}{Theorem}

\newtheorem{remark}{Remark}
\newtheorem{problem}{Problem}

\newcommand{\thmref}[1]{Theorem~\ref{#1}}

\newcommand{\figref}[1]{Fig.~\ref{#1}}

\newcommand{\secref}[1]{Section~\ref{#1}}

\newcommand{\probref}[1]{Problem~\ref{#1}}
\newcommand{\rekref}[1]{Remark~\ref{#1}}

\usepackage{enumerate}

\ifCLASSOPTIONcompsoc
\else
\fi
\ifCLASSINFOpdf
\else
\fi

\graphicspath{{pdfs/},{images/}}

\begin{document}
%
\title{Offline Delay-Optimal Transmission for Energy Harvesting Nodes}
%
%
%
%

\author{\IEEEauthorblockN{Yirui Cong and Xiangyun Zhou}
\IEEEauthorblockA{Research School of Engineering, The Australian National University, Canberra, ACT 0200, Australia\\
Emails: \{yirui.cong, xiangyun.zhou\}@anu.edu.au}
}

\IEEEtitleabstractindextext{%
\begin{abstract}

This paper investigates the offline packet-delay-minimization problem for an energy harvesting transmitter. To overcome the non-convexity of the problem, we propose a C2-diffeomorphic transformation and provide the necessary and sufficient condition for the transformed problem to a standard convex optimization problem. Based on this condition, a simple choice of the transformation is determined which allows an analytically tractable solution of the original non-convex problem to be easily obtained once the transformed convex problem is solved. We further study the structure of the optimal transmission policy in a special case and find it to follow a weighted-directional-water-filling structure. In particular, the optimal policy tends to allocate more power in earlier time slots and less power in later time slots. Our analytical insight is verified by simulation results.

\end{abstract}

}

\maketitle

\IEEEdisplaynontitleabstractindextext

%
\IEEEpeerreviewmaketitle

\section{Introduction}\label{sec:Introduction}

\subsection{Motivation and Related Work}\label{sec:Motivation and Related Work}

Energy harvesting is recognized as a key enabling technology for self-sustainable communication networks. It also brings in significant and necessary changes in the design of the communication protocol when energy harvesting becomes the main or sole source of energy supply for the communication nodes~\cite{GunduzD2014,UlukusS2015,OzelO2015}. By focusing on the design of an energy harvesting transmitter, the existing studies can be categorized into offline and online scenarios.
For offline scenarios, all future information (such as the energy arrival process, data arrival process and channel fading process) is assumed to be predictable or completely deterministic. Hence, the transmission policy is designed offline by using the available knowledge about the future.
For online scenarios, although the exact future behaviors of the energy arrival, data arrival and channel fading processes are not known, the existing literature have considered different levels of statistical knowledge: the statistics of the processes in the future are exactly known (e.g.,~\cite{SharmaV2010}), partly known (e.g.,~\cite{BlascoP2013}) or completely unknown (e.g.,~\cite{CongY2015Submitted1}).

The importance of studying the offline design for energy harvesting transmitters is two-fold~\cite{VazeR2014,OzelO2011,OrhanO2014}: (i) The offline design tells the fundamental performance limit of an energy harvesting communication system. It serves as a benchmark for any online algorithm. Methods such as competitive analysis~\cite{VazeR2014} can be used to derive and analyze online algorithms once the performance of the offline design is known. (ii) The offline solution often helps one to gain insights on the design problem and the behavior of the optimal transmission policy, which inspire possible online design solutions (e.g., Section VI.B in~\cite{OzelO2011} and Section VII in~\cite{OrhanO2014}). As a result, a significant effort has been made into the study of offline designs in the past few years.

The objective of energy harvesting transmission design has been defined from a range of perspectives, including throughput maximization (e.g.,~\cite{OzelO2011,BlascoP2013,OrhanO2014,VazeR2014,CongY2015Submitted1}), remaining energy maximization (e.g.,~\cite{OrhanO2014}), completion time minimization (e.g.,~\cite{YangJ2012TC,OzelO2011,OzelO2012,TutuncuogluK2012TWC,OrhanO2014,CongY2015Submitted1}), and packet delay minimization~\cite{SharmaV2010,ZhangF2014,TongT2015}. In particular, efficient offline designs have been found for the throughput-maximization, remaining-energy-maximization and completion-time-minimization problems~\cite{OzelO2011,OrhanO2014,YangJ2012TC}. However, effective solutions to the offline packet-delay-minimization problem are still yet to be found. A very recent study in~\cite{TongT2015} considered the packet-delay-minimization problem in a non-fading channel and obtained a deep insight of the solution structure given through the KKT conditions. However, the optimal solution was derived for the dual problem rather than the original problem. Since the original problem is non-convex, there is a duality gap~\cite{BoydS2004BOOK} between the original and dual problems.


Another interesting issue is the relationship among the pa-cket-delay-minimization, completion-time-minimization, and throughput-maximization problems.
For the last two problems, they are linked through the maximum departure curve (see Section~V in~\cite{OzelO2011}).
However, it is unclear whether the first problem has any correlation with the last two problems.
%
%
%
Intuitively, the first two problems are related: the quicker the transmission is completed, the smaller the packet delay is.
A similar observation was also made in the literature (e.g.,~\cite{YangJ2010,OrhanO2014}).
%
%
%
Then, through the maximum departure curve, all these three problems appear to be highly related.
%
%
Nevertheless, there has not been any analytical studied on the extent to which the three problems are related and, more importantly, whether the optimal offline transmission designs for the three problems have the similar behaviors.

\subsection{Our Contribution}

In this work, we tackle the aforementioned challenge of finding an effective solution to the offline packet-delay-minimization problem in a general fading channel. We consider a time-slotted system and use the average data queue length to measure delay, which is a commonly adopted delay metric in the literature~\cite{SharmaV2010,ZhangF2014}. Although the formulated delay-minimization problem is non-convex, we are able to transform it into a single convex optimization problem, which can be solved very efficiently.

In order to obtain insights into the behavior of the optimal offline delay-minimizing transmission policy, we focus on a special case where the battery energy is insufficient to clear the data queue in all time slots. In this case, we analytically show that the optimal transmission policy has a weighted-directional-water-filling structure for the power allocation. In particular, the optimal policy tends to allocate more power in earlier time slots and less power in later time slots. This result is further verified by simulations in more general cases.

The insights obtained from the optimal transmission policy also allow us to investigate the relationship and difference among the delay-minimization problem studied in this work, the completion-time minimization problem in~\cite{YangJ2012TC}, and the throughput-maximization problem in~\cite{OzelO2011}.
Surprisingly, the optimal offline solution to the delay-minimization problem is found to be fundamentally different: the delay-minimizing solution tends to reduce the transmit power over time, while the completion-time-minimizing and throughput-maximizing solutions trend to increase the transmit power over time.

\subsection{Paper Organization and Notation}

The system model and problem description are given in \secref{sec:System Model} and \secref{sec:Problem Description: Minimizing the Average Data Queue Length}, respectively.
In \secref{sec:Optimal Solution for the Average-Data-Queue-Minimization Problem}, the delay-minimization problem is efficiently solved and the structure of the optimal solution is studied.
Simulation results are presented in \secref{sec:Simulation Examples} to corroborate our analysis. Finally, the concluding remarks are given in \secref{sec:Conclusion}.
%
%

\section{System Model}\label{sec:System Model}

\begin{figure}[h]
\centering
\includegraphics [width=0.8\columnwidth]{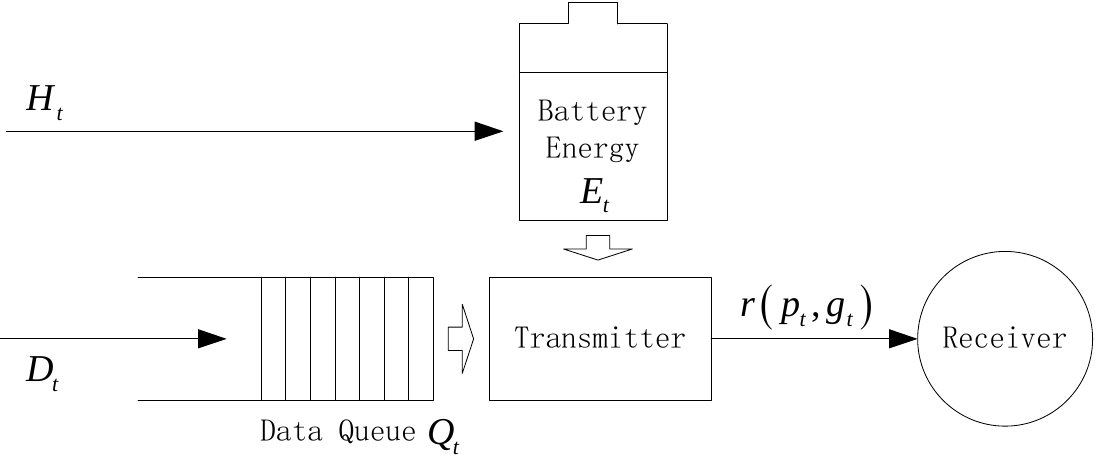}
\caption{The point-to-point communication system with an energy harvesting transmitter.
}
\label{fig:The Transmitter-Receiver Pair}
\end{figure}

We consider a point-to-point communication system with an energy harvesting transmitter (see \figref{fig:The Transmitter-Receiver Pair}).
The time is slotted, and $t \in \{1,\ldots,T\} =: \mathcal{T}$ stands for the $t$\textsuperscript{th} time slot, where $T \in \mathbb{Z}_+$ is the total number of slots.
For each time slot, the battery energy of the transmitter is $E_t \in \overline{\mathbb{R}}_+$, and is determined by the battery energy equation:
\begin{align}\label{eqn:Battery Energy Equation}
E_t = E_{t-1} + H_t - p_t, \quad t \in \mathcal{T},
\end{align}
where $H_t \in \overline{\mathbb{R}}_+$ is the harvested energy and $p_t \in \overline{\mathbb{R}}_+$ is the transmit power.
Note that $E_0 \in \overline{\mathbb{R}}_+$ represents the initial battery energy.
The data queue length of the transmitter is $Q_t \in \overline{\mathbb{R}}_+$, determined by the data queue equation:
\begin{align}\label{eqn:Data Queue Equation}
Q_t = Q_{t-1} + D_t - r(p_t,g_t), \quad t \in \mathcal{T},
\end{align}
where $D_t \in \overline{\mathbb{R}}_+$ is the amount of data arrived and $r(p_t,g_t)$ is the rate function, which is only dependent on the transmit power $p_t$ and the power gain of the communication channel $g_t \in \overline{\mathbb{R}}_+$.
We label $r(p_t,g_t) \equiv r_{g_t}(p_t)$ and assume that $r_{g_t}: \overline{\mathbb{R}}_+ \rightarrow \overline{\mathbb{R}}_+$ is strictly increasing and concave, and has a continuous second derivative, e.g., for fading channel with Gaussian white noise $r_{g_t}(p_t) = \log(1+g_t p_t)$.
Also, $Q_0 \in \overline{\mathbb{R}}_+$ is the initial data queue length.

Note that $E_t$ and $Q_t$ for any $t \in \mathcal{T}$ can be totally determined by~\eqref{eqn:Battery Energy Equation},~\eqref{eqn:Data Queue Equation}, $E_0$ and $Q_0$.
Since we consider an offline scenario, the energy arrival $H_t$, data arrival $D_t$, and channel gain $g_t$ are known in advance.
This completes the description of the system model.

\section{Problem Description: Minimizing the Average Data Queue Length}\label{sec:Problem Description: Minimizing the Average Data Queue Length}

In this paper, we use the average data queue length as the delay metric, which is commonly adopted in the literature~\cite{SharmaV2010,ZhangF2014}.
The average data queue length $\mathcal{L}$ is given by
\begin{align}\label{eqn:Average Data Queue Length}
\mathcal{L} = \frac {1} {T} \sum_{t=1}^{T} Q_t.
\end{align}
%
%
%
%
%
%
%
%
The average-data-queue-minimization problem is to find $p_t$ to minimize $\mathcal{L}$ in~\eqref{eqn:Average Data Queue Length} with constraints from the battery energy equation~\eqref{eqn:Battery Energy Equation} and the data queue length equation~\eqref{eqn:Data Queue Equation}.
Since~\eqref{eqn:Average Data Queue Length} does not explicitly contain $p_t$, we rewrite $\mathcal{L}$ as
\begin{align}\label{eqn:Average Data Queue Length Rewritten}
\begin{split}
\frac {1} {T} \sum_{t=1}^{T} Q_t &\stackrel {(a)}{=} \frac {1} {T} \left\{T Q_0 + \sum_{t=1}^T \sum_{i=1}^t\left[D_i - r_{g_i}(p_i)\right]\right\} \\
&= Q_0 + \frac {1} {T} \sum_{t=1}^{T} \left(T+1-t\right) \left[D_t - r_{g_t}(p_t)\right],
\end{split}
\end{align}
where $(a)$ follows from
\begin{align}
Q_t = Q_0 + \sum_{i=1}^t \left[D_i - r_{g_i}(p_i)\right],~t\in\mathcal{T}.
\end{align}

We also rewrite~\eqref{eqn:Battery Energy Equation} and~\eqref{eqn:Data Queue Equation} to obtain the battery energy and data queue constraints for the optimization problem.
%
%
More specifically, the battery energy equation~\eqref{eqn:Battery Energy Equation} can be rewritten as a series of inequalities, called the battery energy constraints
\begin{align}\label{eqn:Battery Energy Constraints}
E_t = E_0 + \sum_{i=1}^t (H_i - p_i) \geq 0,~t \in \mathcal{T}.
\end{align}
Similarly, we can get the data queue constraints from the data queue equation as
\begin{align}\label{eqn:Data Queue Constraints}
Q_t = Q_0 + \sum_{i=1}^t (D_i - r_{g_i}(p_i)) \geq 0,~t \in \mathcal{T}.
\end{align}
Also note that $p_t$ should be non-negative for $t \in \mathcal{T}$.
To sum up, we can write the offline average-data-queue-minimization problem into an explicit form whose optimization variables are $p_t$ ($t \in \mathcal{T}$), which is given as follows:

\begin{problem}[Average-Data-Queue-Minimization Problem]\label{prob:Average-Data-Queue-Minimization Problem}
The average-data-queue-minimization problem is
\begin{align}\label{eqn:Average-Data-Queue-Minimization Problem}
\begin{array}{c l}
\underset {p_t,\,t\in\mathcal{T}} {\mathrm{minimize}} & Q_0 + \frac {1} {T} \sum_{t=1}^{T} \left(T+1-t\right) \left[D_t - r_{g_t}(p_t)\right]\\
\mathrm{subject~to}~&\sum_{i=1}^{t} p_i \leq E_0 + \sum_{i=1}^{t} H_i,\,t\in\mathcal{T},\\
&\sum_{i=1}^{t} r_{g_i}(p_i) \leq Q_0 + \sum_{i=1}^{t} D_i,\,t\in\mathcal{T},\\
&p_t \geq 0,\,t\in\mathcal{T},
\end{array}
\end{align}
where the optimal solution and optimal objective are labelled as $p_t^*$ ($t \in \mathcal{T}$) and $\mathcal{L}^*$, respectively.
\end{problem}

\begin{remark}
It should be noted that \probref{prob:Average-Data-Queue-Minimization Problem} is not a convex optimization problem, since the data queue constraints correspond to the intersection of non-convex regions w.r.t. $p_t$.
Nevertheless, in \secref{sec:Solving the Average-Data-Queue-Minimization Problem}, we will show that \probref{prob:Average-Data-Queue-Minimization Problem} can be converted to a standard convex optimization problem with transformation on the optimization variables.
\end{remark}

%

\section{Optimal Solution for the Average-Data-Queue-Minimization Problem}\label{sec:Optimal Solution for the Average-Data-Queue-Minimization Problem}

This section is divided into two parts.
In \secref{sec:Solving the Average-Data-Queue-Minimization Problem}, we discuss how to efficiently solve \probref{prob:Average-Data-Queue-Minimization Problem} by transforming it into a standard convex optimization problem.
In \secref{sec:Discussions on the Structure of the Optimal Solution}, we provide insights into the structure of the optimal solution.

\subsection{Solving the Average-Data-Queue-Minimization Problem}\label{sec:Solving the Average-Data-Queue-Minimization Problem}

In this subsection, we discuss how to convert \probref{prob:Average-Data-Queue-Minimization Problem} into a standard convex optimization problem such that it can be solved efficiently.

Let map $\varphi_t : \overline{\mathbb{R}}_+ \rightarrow \overline{\mathbb{R}}_+$ ($t \in \mathcal{T}$) be $C^2$-diffeomorphic\footnote{For each $t \in \mathcal{T}$, $\varphi_t$ is bijective and twice continuously differentiable, and the inverse $\varphi_t^{-1}$ is also twice continuously differentiable.}.
With the transformation $p_t = \varphi_t(q_t)$, \probref{prob:Average-Data-Queue-Minimization Problem} can be rewritten as
\begin{align}\label{eqn:Average-Data-Queue-Minimization Problem Rewritten}
\begin{array}{c l}
\underset {q_t,\,t\in\mathcal{T}} {\mathrm{minimize}} & Q_0 + \frac {1} {T} \sum_{t=1}^{T} \left(T+1-t\right) \left[D_t - r_{g_t}(\varphi_t(q_t))\right]\\
\mathrm{subject~to}~&\sum_{i=1}^{t} \varphi_i(q_i) \leq E_0 + \sum_{i=1}^{t} H_i,\,t\in\mathcal{T},\\
&\sum_{i=1}^{t} r_{g_i}(\varphi_i(q_i)) \leq Q_0 + \sum_{i=1}^{t} D_i,\,t\in\mathcal{T},\\
&\varphi_t(q_t) \geq 0,\,t\in\mathcal{T},
\end{array}
\end{align}
where $r_{g_t}(\varphi_t(q_t))$ is twice continuously differentiable, since both $r_{g_t}$ and $\varphi_t$ are twice continuously differentiable.

The following theorem gives the necessary and sufficient condition for guaranteeing the convexity of~\eqref{eqn:Average-Data-Queue-Minimization Problem Rewritten}.

\begin{theorem}[Necessary and Sufficient Condition for Convexity of Transformed Problem]\label{thm:Necessary and Sufficient Condition for Convexity of Transformed Problem}
Problem~\eqref{eqn:Average-Data-Queue-Minimization Problem Rewritten} is a standard convex optimization problem if and only if $\forall t \in \mathcal{T}$,
\begin{align}\label{eqn:Necessary and Sufficient Condition for Convexity of Transformed Problem}
\varphi_t (q_t) = r_{g_t}^{-1} (a_t q_t + b_t),
\end{align}
where $a_t \in \mathbb{R}_+$ and $b_t \in \mathbb{R}$.
\end{theorem}

\begin{IEEEproof}
The proof is divided into three steps.
In the first step, we show that the condition $\varphi_t(q_t) \geq 0$ is equivalent to $q_t \geq 0$ for all $t \in \mathcal{T}$.
In the second step, we prove the necessary and sufficient condition for the convexity of Problem~\eqref{eqn:Average-Data-Queue-Minimization Problem Rewritten} is that the map $\varphi_t$ is convex and the composition $r_{g_t} \circ \varphi_t$ is affine.
In the third step, we show that~\eqref{eqn:Necessary and Sufficient Condition for Convexity of Transformed Problem} holds.

i)~Since, for each $t \in \mathcal{T}$, $\varphi_t$ is bijective from $\overline{\mathbb{R}}_+$ to $\overline{\mathbb{R}}_+$, we can derive that if $p_t = \varphi_t(q_t) \in \overline{\mathbb{R}}_+$, then $q_t \in \overline{\mathbb{R}}_+$, vice versa.
Thus, the domains of $\varphi_t(q_t) \geq 0$ and $q_t \geq 0$ are totally the same, and Problem~\eqref{eqn:Average-Data-Queue-Minimization Problem Rewritten} is equivalent to the following problem
\begin{align}\label{eqn:Average-Data-Queue-Minimization Problem Rewritten 2}
\begin{array}{c l}
\underset {q_t,\,t\in\mathcal{T}} {\mathrm{minimize}} & Q_0 + \frac {1} {T} \sum_{t=1}^{T} \left(T+1-t\right) \left[D_t - r_{g_t}(\varphi_t(q_t))\right]\\
\mathrm{subject~to}~&\sum_{i=1}^{t} \varphi_i(q_i) \leq E_0 + \sum_{i=1}^{t} H_i,\,t\in\mathcal{T},\\
&\sum_{i=1}^{t} r_{g_i}(\varphi_i(q_i)) \leq Q_0 + \sum_{i=1}^{t} D_i,\,t\in\mathcal{T},\\
&q_t \geq 0,\,t\in\mathcal{T}.
\end{array}
\end{align}

ii)~Now, we prove the necessary and sufficient condition for the convexity of Problem~\eqref{eqn:Average-Data-Queue-Minimization Problem Rewritten 2} instead of Problem~\eqref{eqn:Average-Data-Queue-Minimization Problem Rewritten}.

Necessity.
If Problem~\eqref{eqn:Average-Data-Queue-Minimization Problem Rewritten 2} is a standard convex optimization problem, then the objective function and the inequality constraints are convex (see Chapter 4.2 in~\cite{BoydS2004BOOK}).
Firstly, we focus on the battery energy constraints (the first row of the constraints in~\eqref{eqn:Average-Data-Queue-Minimization Problem Rewritten 2}).
For $t = 1$, we have $\varphi_1(q_1) \leq E_0 + H_1$ and this means that $\varphi_1(q_1)$ is convex w.r.t. $q_1$.
For $t = 2$, we have $\varphi_1(q_1) + \varphi_2(q_2)$ is convex w.r.t. $q_1$ and $q_2$, which implies $\varphi_2(q_2)$ is convex w.r.t. $q_2$, since $\varphi_1(q_1)$ is convex w.r.t. $q_1$.
Likewise, we can derive that $\varphi_t$ is convex for all $t \in \mathcal{T}$.

Similarly, the data queue constraints (the second row of the constraints in~\eqref{eqn:Average-Data-Queue-Minimization Problem Rewritten 2}) tell that $r_{g_t}\circ\varphi_t$ is also convex for all $t \in \mathcal{T}$.

Furthermore, with convex $r_{g_t}\circ\varphi_t$, we see that the objective function is concave, which implies it is affine (both convex and concave), so does the map $r_{g_t}\circ\varphi_t$.
This completes the necessity.

Sufficiency.
If $\varphi_t$ is convex and $r_{g_t}\circ\varphi_t$ is affine (affinity also means convexity) for $t \in \mathcal{T}$, then the objective function as well as all the inequality constraints are convex.
Therefore, Problem~\eqref{eqn:Average-Data-Queue-Minimization Problem Rewritten 2} is a convex optimization problem.

iii)~Since the composition $r_{g_t}\circ\varphi_t$ is affine and invertible, we have $r_{g_t}(\varphi_t(q_t)) = a_t q_t + b_t$, where $a_t \in \mathbb{R}\setminus \{0\}$ and $b_t \in \mathbb{R}$.
This means $\varphi_t(q_t) = r_{g_t}^{-1} (a_t q_t + b_t)$.
Note that $r_{g_t}^{-1}$ is convex, and to make $\varphi_t$ convex, $a_t$ should be greater than $0$.
To sum up, condition~\eqref{eqn:Necessary and Sufficient Condition for Convexity of Transformed Problem} is established.
\end{IEEEproof}

\thmref{thm:Necessary and Sufficient Condition for Convexity of Transformed Problem} gives an important guideline for selecting the transformation $\varphi_t$.
With this guideline, we further choose $a_t = 1$ and $b_t = 0$ for all $t \in \mathcal{T}$ such that the transformation is $\varphi_t = r_{g_t}^{-1}$, which can largely reduce the complexity of the transformed problem.
Therefore, we obtain the following transformed problem by applying the transformation $p_t = r_{g_t}^{-1}(q_t)$ for \probref{prob:Average-Data-Queue-Minimization Problem}.

\begin{problem}[Transformed Problem]\label{prob:Transformed Problem}
With $p_t = r_{g_t}^{-1}(q_t)$ for all $t \in \mathcal{T}$, \probref{prob:Average-Data-Queue-Minimization Problem} is transformed into the following convex optimization problem
\begin{align}\label{eqn:Transformed Problem}
\begin{array}{c l}
\underset {q_t,\,t\in\mathcal{T}} {\mathrm{minimize}} & Q_0 + \frac {1} {T} \sum_{t=1}^{T} \left(T+1-t\right) \left(D_t - q_t\right)\\
\mathrm{subject~to}~&\sum_{i=1}^{t} r_{g_i}^{-1}(q_i) \leq E_0 + \sum_{i=1}^{t} H_i,\,t\in\mathcal{T},\\
&\sum_{i=1}^{t} q_i \leq Q_0 + \sum_{i=1}^{t} D_i,\,t\in\mathcal{T},\\
&q_t \geq 0,\,t\in\mathcal{T},
\end{array}
\end{align}
where the optimal solution is $q_t^*$ ($t \in \mathcal{T}$), but not necessarily unique\footnote{Since the objective of \probref{prob:Transformed Problem} is not \textit{strictly} convex, we cannot ensure that it always has a unique global minimizer under any given $H_t$, $D_t$, $E_0$ and $Q_0$. However, since this problem is convex, we can still get at least one global minimizer using an efficient method, e.g., interior-point method.}.
The optimal objective is the same as that in \probref{prob:Average-Data-Queue-Minimization Problem}, i.e., $\mathcal{L}^*$.
\end{problem}

\begin{remark}[Solving the Original Problem with the Solution of the Transformed Problem]\label{rek:Solving the Original Problem with the Transformed Problem}
%
%
The optimal solution of \probref{prob:Average-Data-Queue-Minimization Problem} can be derived in the following way
\begin{align}\label{eqn:Solving the Original Problem with the Transformed Problem}
p_t^* = r_{g_t}^{-1}(q_t^*),~t \in \mathcal{T},
\end{align}
where $q_t^*$ is the optimal solution to \probref{prob:Transformed Problem}.
This implies that we can use~\eqref{eqn:Solving the Original Problem with the Transformed Problem} to obtain the optimal solution of \probref{prob:Average-Data-Queue-Minimization Problem} after deriving the optimal solution of \probref{prob:Transformed Problem}.
Since \probref{prob:Transformed Problem} is convex, it can be solved efficiently (e.g., using the interior-point method~\cite{BoydS2004BOOK}).
Therefore, the optimal solution of \probref{prob:Average-Data-Queue-Minimization Problem} can also be derived efficiently.
\end{remark}

\subsection{Discussions on the Structure of the Optimal Solution}\label{sec:Discussions on the Structure of the Optimal Solution}

In general, it is difficult to derive the structure of the optimal solution of \probref{prob:Transformed Problem} by its KKT conditions.
However, the KKT conditions can give insights into the structure of the optimal solution when $Q_t > 0$ for all $t \in \mathcal{T}$.
Such a case happens when the transmitter does not have sufficient energy in the battery to completely clear the data queue in all time slots.
In order to shed light on the structure of the delay-minimization transmission policy, we study this special case using the original problem (\probref{prob:Average-Data-Queue-Minimization Problem}) rather than \probref{prob:Transformed Problem}.

In this special case, since $Q_t > 0$ for all $t \in \mathcal{T}$, the data queue constraints in \probref{prob:Average-Data-Queue-Minimization Problem} can be removed.
We further assume the channel is a fading channel with Gaussian white noise, and the rate function is $r_{g_t}(p_t) = \log (1+g_t p_t)$, which achieves the channel capacity.
Noting that to minimize $Q_0 + \frac {1} {T} \sum_{t=1}^{T} (T+1-t) [D_t - r_{g_t}(p_t)]$ is equivalent to maximize $\sum_{t=1}^{T} (T+1-t) r_{g_t}(p_t)$, we can rewrite \probref{prob:Average-Data-Queue-Minimization Problem} as the following problem.

\begin{problem}[Weighted-Throughput-Maximization Problem]\label{prob:Weighted-Throughput-Maximization Problem}
%
\begin{align}\label{eqn:Weighted-Throughput-Maximization Problem}
\begin{array}{l c l}
&\underset {p_t,\,t\in\mathcal{T}} {\mathrm{maximize}} & \sum_{t=1}^{T} \left(T+1-t\right) \log (1+g_t p_t)\\
&\mathrm{subject~to}~&\sum_{i=1}^{t} p_i \leq E_0 + \sum_{i=1}^{t} H_i,\,t\in\mathcal{T},\\
&~&p_t \geq 0,\,t\in\mathcal{T}.
\end{array}
\end{align}
\end{problem}

\begin{figure*}
\centering
\subfigure[]{\includegraphics [width=0.6\columnwidth]{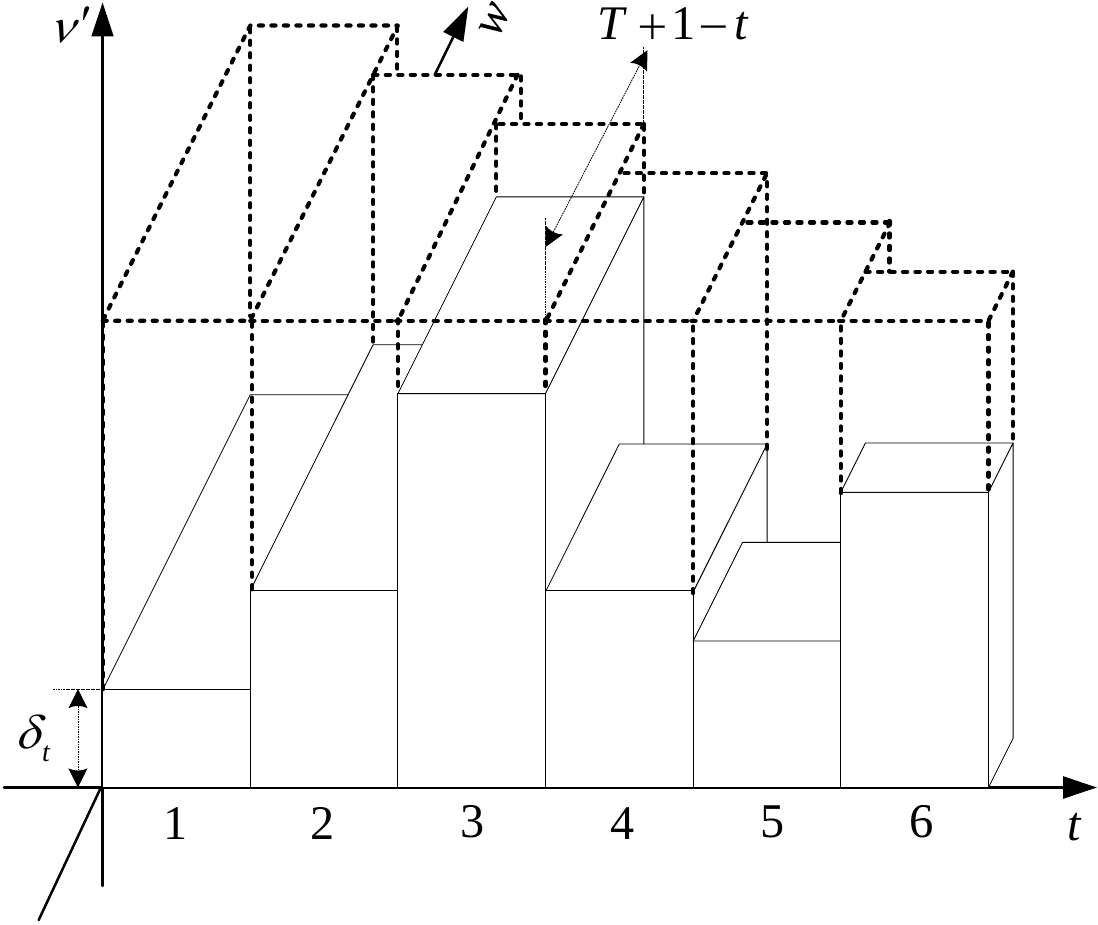}\label{fig:Directional Water Filling Algorithm A}}\hspace{0.1em}
\subfigure[]{\includegraphics [width=0.6\columnwidth]{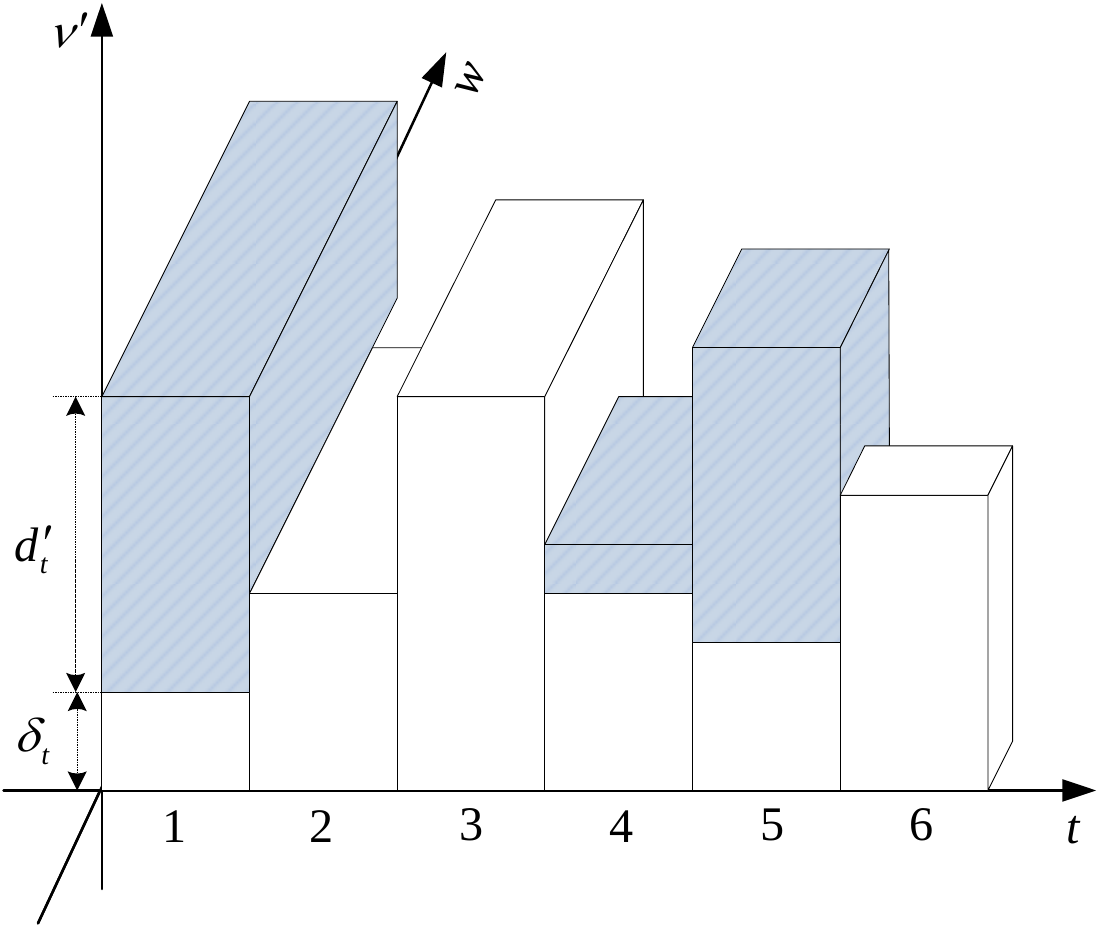}\label{fig:Directional Water Filling Algorithm B}}\hspace{0.1em}
\subfigure[]{\includegraphics [width=0.6\columnwidth]{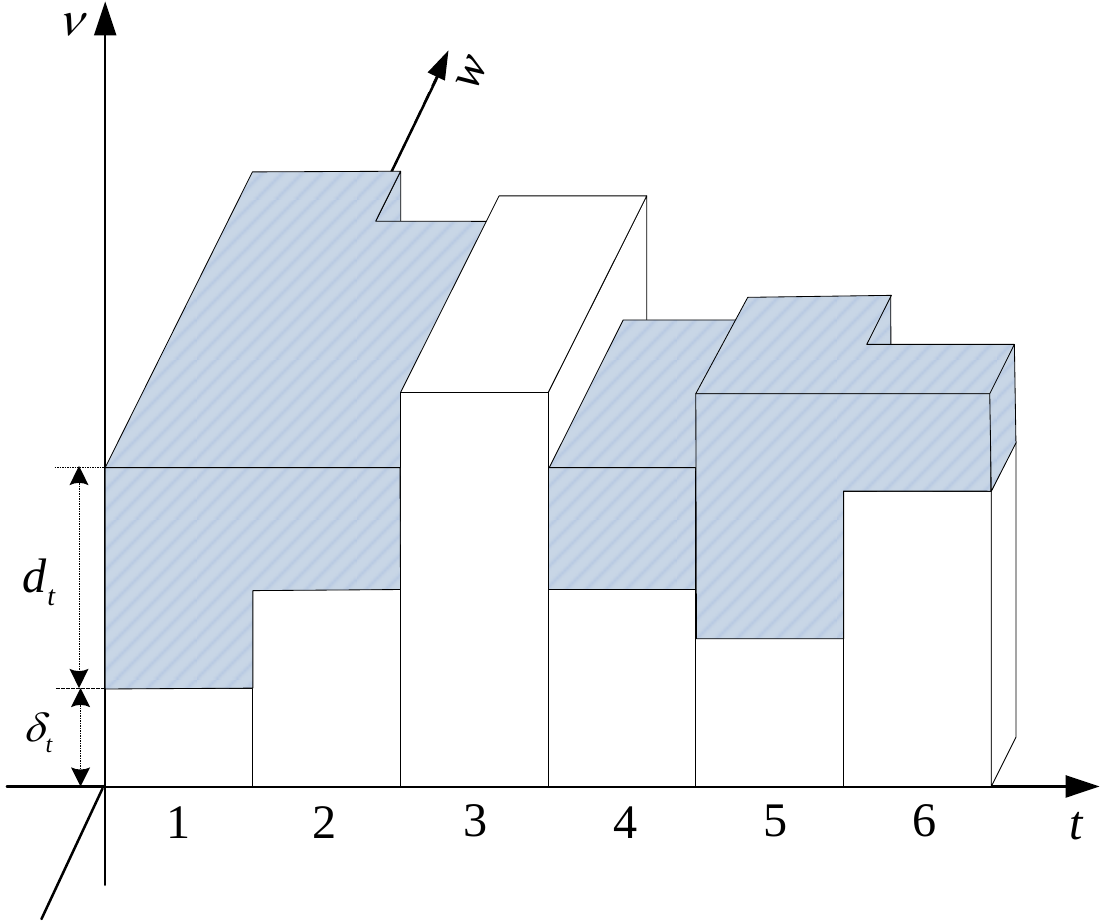}\label{fig:Directional Water Filling Algorithm C}}

\caption{An illustration of the weighted-directional-water-filling algorithm.
The $t$-axis refers to the time slot, and the $w$-axis stands for the weight $T+1-t$.
(a) The shape of the water tank and the levels of the ground.
(b) The original water levels, which are determined by the initial battery energy and energy arrival process.
(c) The new water levels after applying the weighted-directional-water-filling algorithm.
}\label{fig:Directional Water Filling Algorithm}
\end{figure*}

Interestingly, \probref{prob:Weighted-Throughput-Maximization Problem} maximizes the throughput over all time slots with a weighting factor $(T+1-t)$ applied to each time slot.
It is a convex optimization problem and \textit{has a unique maximizer}, because $\log (1+g_t p_t)$ is strictly concave for all $t \in \mathcal{T}$ and the constraints are all affine.
We claim that the solution to \probref{prob:Weighted-Throughput-Maximization Problem} has a weighted-directional-water-filling structure\footnote{Note that the mathematical form of \probref{prob:Weighted-Throughput-Maximization Problem} is similar to~(38a) and~(38b) in~\cite{TutuncuogluK2015}, where the weights are preferences to broadcast channels.
However, our solution has a more explanation (see \figref{fig:Directional Water Filling Algorithm}).}, which is an extension of the directional-water-filling structure in~\cite{OzelO2011}.
%

Applying KKT conditions to \probref{prob:Average-Data-Queue-Minimization Problem}, we can derive the structure of the optimal power as
\begin{align}\label{eqn:Optimal Powers}
p_t^* = \left(\frac {T+1-t}{\sum_{i=t}^{T}\lambda_i - \eta_t} - \frac {1}{g_t}\right)^+,
\end{align}
where $\lambda_t$ and $\eta_t$ correspond to the battery energy constraints (the first row of the constraints in~\eqref{eqn:Weighted-Throughput-Maximization Problem}) and $p_t \geq 0$.
Defining
\begin{align}\label{eqn:Water Levels}
\nu_t = \frac {1} {\sum_{i=t}^{T}\lambda_i - \eta_t},
\end{align}
we have the following theorem.

\begin{theorem}[Nondecreasing Property of Water Levels]\label{thm:Nondecreasing Property of Water Levels}
Let the time set $\mathcal{T}_0$ contains those $t$ corresponding to $p_t^* = 0$, i.e., $\mathcal{T}_0 = \{t: p_t^* = 0, t \in \mathcal{T}\}$.
We have $\nu_{t+1} \geq \nu_t$ for all $t \in \{1,\ldots,T-1\}\backslash{\mathcal{T}_0}$.
Furthermore, $\forall t \in \{1,\ldots,T-1\}\backslash{\mathcal{T}_0}$, if the battery energy constraint (the first row of the constraints in~\eqref{eqn:Weighted-Throughput-Maximization Problem}) is satisfied without equality, then $\nu_{t+1} = \nu_t$.
\end{theorem}

\begin{IEEEproof}
When $p_t^* > 0$, it follows the complementary slackness (see Chapter 5.5.2 in~\cite{BoydS2004BOOK}) that $\eta_t = 0$, and $\nu_t$ becomes $\nu_t = {1} / {\sum_{i=t}^{T}\lambda_i}$.
Observing that $\lambda_t \geq 0$, we have $\nu_{t+1} \geq \nu_t$ for all $t \in \{1,\ldots,T-1\}\backslash{\mathcal{T}_0}$.
Moreover, $\forall t \in \{1,\ldots,T-1\}\backslash{\mathcal{T}_0}$, if the battery energy constraint (the first row of the constraints in~\eqref{eqn:Weighted-Throughput-Maximization Problem}) is satisfied without equality, then $\lambda_t = 0$, which implies $\nu_{t+1} = \nu_t$.
\end{IEEEproof}

\thmref{thm:Nondecreasing Property of Water Levels} tells that the optimal solution $p_t^*$ has the weighted-directional-water-filling structure, and $\nu_t$ is exactly the water level in the $t$\textsuperscript{th} time slot.
The water depth $d_t$ is
\begin{align}\label{eqn:Water Depths}
d_t := \frac {p_t^*} {T+1-t} = \left(\nu_t - \delta_t\right)^+,~t \in \mathcal{T}
\end{align}
where $\delta_t = 1/[(T+1-t)g_t]$.
Therefore, the weighted-directional-water-filling algorithm can be derived, which is illustrated in \figref{fig:Directional Water Filling Algorithm}.

In \figref{fig:Directional Water Filling Algorithm A}, the white blocks at the bottom are the ground with levels $\delta_t = 1/[(T+1-t)g_t]$, corresponding to the effect of the channel gain.
The above dash lines complement the shape of the water tank, where the widths $(T+1-t)$ are decreasing linearly with $t$.
The walls between any adjacent time slots are right permeable~\cite{OzelO2011}, which means the water can flow to the right if the water in the left slot has a higher level, but the water cannot flow back to the left in all cases.
In \figref{fig:Directional Water Filling Algorithm B}, the shaded blocks are water, which is determined by $E_0$ as well as $H_t$ arriving in each time slot.
In this particular example, the water depths in time slots $1$, $4$, and $5$ are $d'_1 = (E_0 + H_1)/T$, $d'_4 = H_4/T$ and $d'_5 = H_5/T$, respectively.
Thus, the water levels in these three time slots can be calculated by $\nu'_t = \delta_t + d'_t$.
On the other hand, the energy arrived is zero in time slots $2$, $3$, and $6$, and hence, no water exists in those time slots.
%
%
In \figref{fig:Directional Water Filling Algorithm C}, the water is filled across time slots, and the final water volume for each time slot is exactly the power allocation.
The new water level, after using the weighted-directional-water-filling algorithm, is $\nu_t = \delta_t + d_t$ for all $t \in \mathcal{T}$.
In this example, time slots $1$, $2$ and $4$ share a same water level, since the original water level $\nu'_1$ in slot $1$ is higher than $\nu'_2$ and $\nu'_4$, so is the water level in time slots $5$ and $6$.
However, there is no water pouring from time slot $5$ to $4$, even though $\nu'_4 < \nu'_5$, because the walls are right permeable.
This completes the description of the weighted-directional-water-filling algorithm in this example.

\begin{remark}\label{rek:Decreasing Tendency of Allocated Powers}
From \figref{fig:Directional Water Filling Algorithm}, we observe that the weight $(T+1-t)$ is initially large and decreases as time goes by.
As a result, the optimal solution of \probref{prob:Weighted-Throughput-Maximization Problem} tends to allocate more transmit power in the earlier time slots, which implies the transmit power is more likely to decrease with $t$.
Furthermore, if $H_t = 0$ for $t \in \{2,\ldots,T\}$ and $g_t \equiv 1$ for $t \in \mathcal{T}$, we can prove that the transmit power $p_t$ is linearly decreasing with $t$.
%
%
%
\end{remark}

As mentioned in \secref{sec:Motivation and Related Work}, the delay-minimization problem studied in this work was previously considered to be highly related to the completion-time-minimization problem, and the throughput-maximization problem.
However, \rekref{rek:Decreasing Tendency of Allocated Powers} tells a fact that the completion-time-minimization and throughput-maximization problems are inherently different from the delay-minimization problem.
This is because the allocated transmit power in the solutions to the completion-time-minimization and throughput-maximization problems (if the battery storage is infinite and the channel is non-fading) are increasing with time $t$ (see~\cite{YangJ2012TC} and Theorem~1 in~\cite{OzelO2011}), while the optimal power allocation in the delay-minimization problem tends to decrease with $t$.


\section{Simulation Results}\label{sec:Simulation Examples}

In this section, firstly, we numerically illustrate that the optimal solution to the delay-minimization problem is fundamentally different from that to the completion-time-minimization and throughput-maximization problems.
Note that the simulations contain more cases than that in \secref{sec:Discussions on the Structure of the Optimal Solution}, e.g., not limited to the case that the transmitter has insufficient energy in the battery to clear the data queue. The Monte Carlo method is employed to investigate the behavior of the optimal solution to \probref{prob:Average-Data-Queue-Minimization Problem}, which can be efficiently solved by our results in \secref{sec:Solving the Average-Data-Queue-Minimization Problem}.
Secondly, we compare the packet delay between our delay-minimization algorithm and the throughput-maximization algorithm to stress the optimality of our algorithm.

To focus on the decreasing or increasing property of the power allocation over time, we introduce a concept, named the inversion number.
The inversion number for a sequence of power values $(p_t)_{t\in\mathcal{T}}$ is defined as
\begin{align}\label{eqn:Inversion Number}
\mathrm{inv}((p_t)_{t\in\mathcal{T}}) := \# \left\{\left(p_{t_1}, p_{t_2}\right): t_1 < t_2 \wedge p_{t_1} > p_{t_2}\right\},
\end{align}
where $\#$ returns the cardinality of a finite set, and $\wedge$ means the logical ``and''.
It can be observed that: if a sequence is non-decreasing, then the inversion number is $0$;
while, if a sequence is strictly decreasing, the inversion number is $T(T+1)/2$.
Thus, the larger the inversion number is, the more likely a sequence $(p_t)_{t\in\mathcal{T}}$ is decreasing with $t$.

\begin{figure}[h]
\centering
\includegraphics [width=0.8\columnwidth]{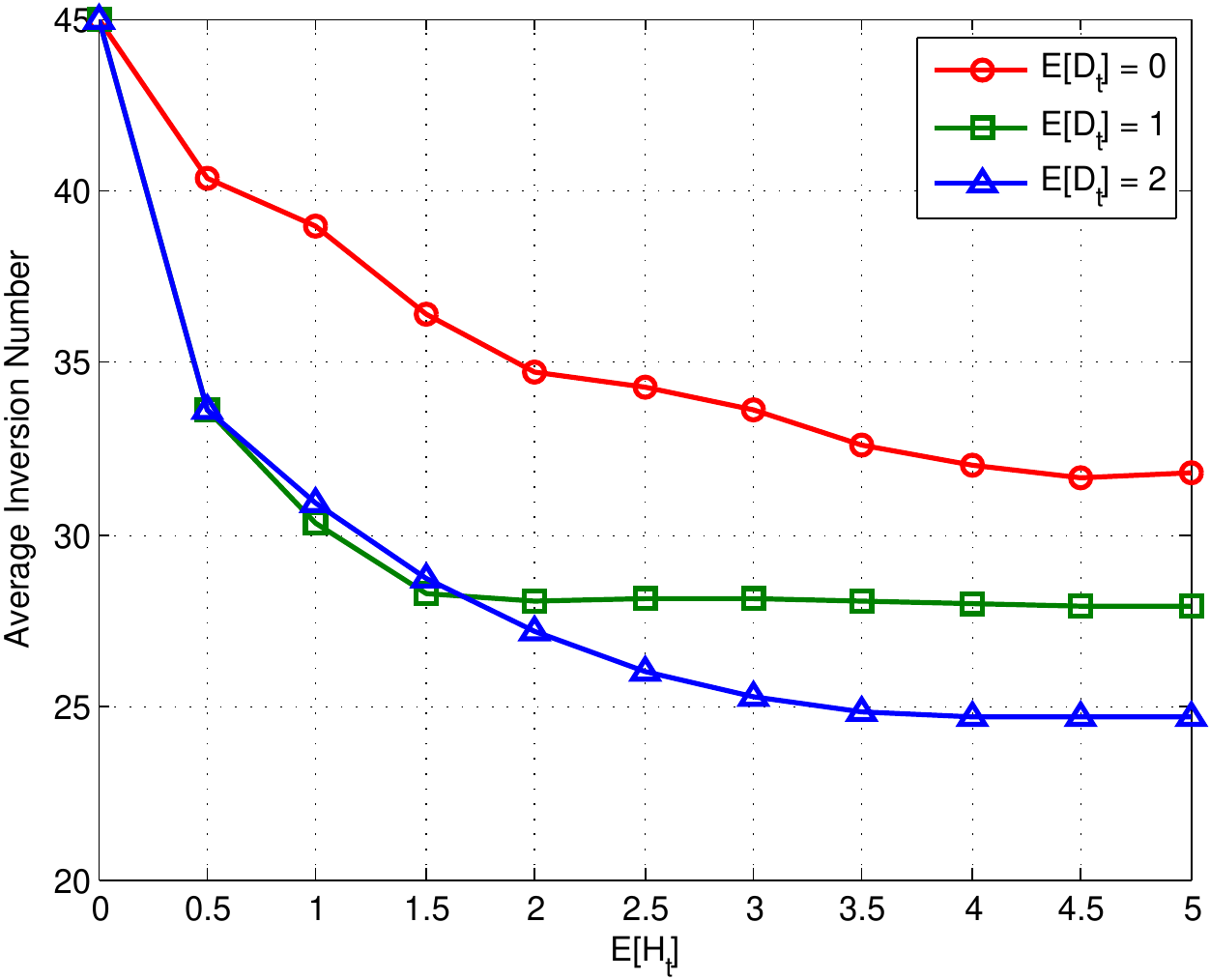}
\caption{Average inversion numbers.
}
\label{fig:Average Inversion Number of Powers}
\end{figure}

The simulation parameters are given as follows.
The number of time slots is $T = 10$, i.e., the simulations are done within the time horizon $\{1,\ldots,10\}$.
The initial battery energy and the initial data queue length are $E_0 = 1$ and $Q_0 = 1$, respectively.
We set the power gain of the channel as $g_t \equiv 1$, with the rate function given by $\log(1 + g_t p_t)$, in order to compare with the power allocation solution to the completion-time-minimization problem derived in~\cite{YangJ2012TC}.
The energy arrival process $H_t$ follows an independent and identically distributed (i.i.d.) uniform distribution, so does the data arrival process $D_t$.
Additionally, the mean values $\mathbb{E}[H_t]$ and $\mathbb{E}[D_t]$ are set within $\{0,0.5,\ldots,5\}$ and $\{0,1,2\}$, respectively (see \figref{fig:Average Inversion Number of Powers}).
Each point in \figref{fig:Average Inversion Number of Powers} is obtained from averaging over $10,000$ simulation runs.

From \figref{fig:Average Inversion Number of Powers}, we have the following observations:
For a fixed $\mathbb{E}[D_t]$, the average inversion number decreases with $\mathbb{E}[H_t]$.
More importantly, the case $\mathbb{E}[H_t] = 0$ corresponds to a strictly decreasing sequence of power values, since the average inversion number is exactly $T(T+1)/2 = 45$.
Note that, in this case ($\mathbb{E}[H_t] = 0$ means $H_t \equiv 0$), the decreasing property has been explained analytically in \rekref{rek:Decreasing Tendency of Allocated Powers}.
When $\mathbb{E}[H_t]$ goes large enough, the average inversion number for a fixed $\mathbb{E}[D_t]$ gradually converges to a constant.
This is because, in every time slot, there is sufficient battery energy to clear the data queue with high probability, and hence, the sequence of power values is largely dependent on $D_t$, and independent of $H_t$.
We can see that for $\mathbb{E}[D_t] \in \{0, 1, 2\}$, the average inversion numbers converge to $31.8$, $27.9$, and $24.7$, respectively, which are far from $0$.
According to the results in~\cite{YangJ2012TC} and~\cite{OzelO2011}, the sequences of power values of the completion-time-minimization and the throughput-maximization solutions are non-decreasing, hence their inversion numbers are $0$.
This confirms that the optimal solution of the delay-minimization problem has a very different behavior from those of the completion-time-minimization and throughput-maximization problems.

\begin{figure}[h]
\centering
\includegraphics [width=0.8\columnwidth]{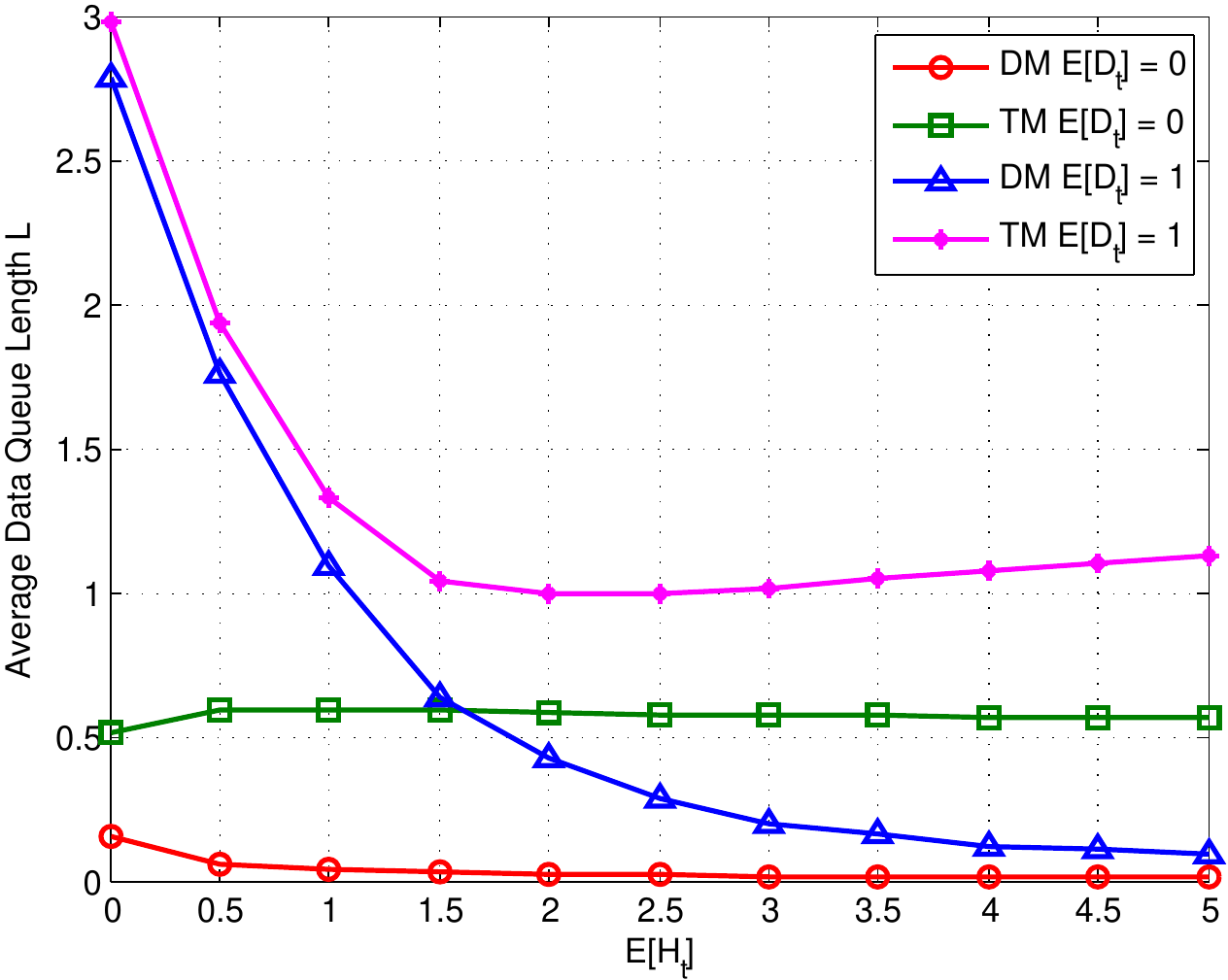}
\caption{Packet delay comparisons between the delay-minimization algorithm (labeled by DM) and the throughput-maximization algorithm (labeled as TM).
}
\label{fig:System Delays}
\end{figure}

Using the same simulation parameters but with the channel power gain $g_t$ following i.i.d. Nakagami-$2$ fading, we compare our delay-minimization algorithm with the throughput-maximization algorithm (adopted from~\cite{OzelO2011}).
The results are shown in \figref{fig:System Delays}, and we can see that the delay-minimization algorithm performs much better, and the throughput-maximization algorithm does not give good delay performance.
This result again confirms the fundamental difference between the delay-minimization and throughput-maximization problems.
%
%

\section{Conclusion}\label{sec:Conclusion}

The offline packet-delay-minimization problem for an energy harvesting transmitter has been investigated in a time-slotted system. We have proposed an effective transformation that converts the original non-convex problem into a convex problem, which then can be solved very efficiently. We have found that the optimal offline transmission policy of
the packet-delay-minimization design is fundamentally different from those of the completion-time-minimization and throughput-maximization designs reported in the literature. The former one tends to decrease the allocated power as time increases, while the latter two allocate more power as time increases.
%

%

\bibliographystyle{IEEEtran}

\bibliography{OfflineDelayOptimization}
\end{document}